\documentclass[10pt,journal]{IEEEtran}
\usepackage{cite}
\ifCLASSINFOpdf
\else
  \usepackage[dvips]{graphicx}
  \DeclareGraphicsExtensions{.eps}
\fi
\usepackage[cmex10]{amsmath}
\usepackage{amsfonts,amssymb,amsthm,mdwmath}
\newtheorem{remark}{Remark}

\usepackage{array}

\usepackage{stfloats}
\usepackage{url}
\usepackage{acronym}
\begin{document}
\acrodef{LoS}[LoS]{Line-of-Sight}
\acrodef{NLoS}[NLoS]{Non-Line-of-Sight}
\acrodef{PDF}[PDF]{Probability Distribution Function}
\acrodef{RV}[RV]{Random Variable}
%\acrodef{PPP}[PPP]{Poisson Point Process}
\acrodef{MGF}[MGF]{Moment Generating Function}
\acrodef{MAC}[MAC]{Medium Access Control}
\acrodef{RWPM}[RWPM]{Random Waypoint Mobility}
\acrodef{i.i.d.}[i.i.d.]{independent and identically distributed}
\acrodef{mmW}[mmW]{millimeter-wave wireless}
\acrodef{1D}[1D]{one-dimensional}

\title{Correlated Interference from Uncorrelated Users in Bounded Ad Hoc Networks with Blockage}
%Scaling Limits for Temporal Correlation of Interference in Ad Hoc Networks with Blockage

%\author{Author 1 and Author 2 % <-this % stops a space
% \IEEEcompsocitemizethanks{\IEEEcompsocthanksitem The authors are with the ... \protect \\
%% % \\ is fragile and will error, could use \hfil\break instead.
%% E-mail: \{K.Koufos, Carl.Dettmann\}@bristol.ac.uk }
%% %\IEEEcompsocthanksitem J. Doe and J. Doe are with Anonymous University.}% <-this % stops an unwanted space
%% %\thanks{Manuscript received April 19, 2005; revised August 26, 2015.} 
%% 
%}}

%\IEEEcompsocitemizethanks{\IEEEcompsocthanksitem

\author{Konstantinos Koufos, Carl P. Dettmann and Justin P. Coon % <-this % stops a space
\thanks{K.~Koufos and C.P.~Dettmann are with the School of Mathematics, University of Bristol, BS8 1TW, Bristol, UK. \{K.Koufos, Carl.Dettmann\}@bristol.ac.uk} \protect \\ 
\thanks{J.P.~Coon is with the Department of Engineering Science, University of Oxford, Oxford, OX1 3PJ, UK. justin.coon@eng.ox.ac.uk} }

\maketitle

\begin{abstract}
In this letter, we study the joint impact of user density, blockage density and deployment area on the temporal correlation of interference for static $\left(u\!=\!0\right)$ and highly mobile $\left(u\!\rightarrow\!\infty\right)$ users. We show that even if the user locations become uncorrelated in the limit of $u\!\rightarrow\!\infty$, the interference level can still be correlated when the deployment area is bounded and/or there is blockage. In addition, we study how the correlation coefficients of interference scale at a high density of blockage. 
%We show that the correlation coefficient does not vanish to zero when the deployment area is bounded. With infinite boundaries, the correlation coefficient can stay  positive when there is correlated slow fading due to blockage. We illustrate the joint impact of user and blockage densities on the correlation coefficients and study how the coefficients scale at a high blockage density. 
\end{abstract}


% \fi
\IEEEpeerreviewmaketitle

\section{Introduction}
\label{sec:Introduction}

\IEEEPARstart{T}{he} correlation of interference over sequential periods of time is an important quantity to study because it affects the correlation of receiver outage, the end-to-end delay, the handoff rate etc.~\cite{Gong2014, Dhillon2016}. It arises due to correlations in the propagation channel and the \ac{MAC} scheme~\cite{Gong2014,Schilcher2012}. For ALOHA type of MAC, the propagation conditions become correlated in time when there are correlations in the fading channel and the user mobility. 

Keeping in mind the ongoing standardization activities for the deployment of commercial \ac{mmW} networks, the impact of blockage and deployment area on the correlation of interference becomes an attractive topic to study. Thus far, the performance analysis of static \ac{mmW} networks, e.g.~\cite{Bai2015,Thornburg2016} neglects the correlation of links that  share common obstacles and also assumes infinite deployment area. In~\cite{Koufos2016b}, these assumptions are not adopted, however, only the  short-term correlation of interference is studied using the \ac{RWPM} model, and the user locations are discrete.

%In this paper, we study the temporal correlation of interference for: (i) static users $\left(u\!=\!0\right)$, and (ii) users with \ac{i.i.d.} locations over time. The former is useful for studying static networks. The latter can be used to calculate the correlation of interference in highly moving networks $\left(u\!\rightarrow\!\infty\right)$ and/or the long-term interference correlation in networks with asymptotic independent mobility, e.g., random walk, Brownian motion, constrained \ac{i.i.d.} mobility with wrap around or bouncing back~\cite{Gong2014, Bettstetter2001}, \ac{RWPM} ~\cite{Paolo2003}, etc. Studying interference correlation with uncorrelated mobility will also highlight that a bounded domain and/or a domain with blockage can make the interference pattern correlated too.

In this paper, we consider a continuous and bounded \ac{1D} deployment, and we study the temporal correlation of interference for static users $\left(u\!=\!0\right)$, and users with \ac{i.i.d.} locations over time. The former is useful for studying static networks. The latter can be used to calculate the correlation of interference in highly moving networks $\left(u\!\rightarrow\!\infty\right)$ and/or the long-term interference correlation in networks with asymptotic independent mobility, e.g., random walk, Brownian motion, constrained \ac{i.i.d.} mobility with wrap around or bouncing back~\cite{Gong2014, Bettstetter2001}, \ac{RWPM} ~\cite{Paolo2003}, etc. Studying interference correlation with uncorrelated mobility will also highlight that a bounded domain and/or a domain with blockage can make the interference pattern correlated too. Even though the analysis in the \ac{1D} space seems to be an over-simplification, it allows getting useful insights about the correlation of interference at a low complexity. The \ac{1D} scenario can also find practical applications, e.g., in vehicular networks. 
%We consider both static networks, $\left(u\!=\!0\right)$, and a more general class of mobile networks as compared to~\cite{Koufos2016b}, i.e., networks with a high user speed, $\left(u\!\rightarrow\!\infty\right)$, and/or asymptotic independent mobility.

Next, we summarize the most important  insights about the system behaviour which, to the best of our knowledge, are new: (i) With uncorrelated user mobility, the temporal correlation of interference becomes inversely proportional to the size of the deployment domain when there is no blockage. (ii) With a finite density of blockage, the correlation coefficient stays  positive, even if the deployment area is infinite. This is because blockage introduces correlation in the interference levels generated by different users.
(iii) In the static case, blockage increases the correlation of interference. (iv) With uncorrelated mobility, there is a critical user-to-blockage density ratio that determines the correlation of interference as compared to the case without blockage. At a high density of blockage, the critical ratio can be expressed in a closed-form.

\section{System model}
\label{sec:System}
We consider two independent Poisson Point Processes (PPPs), one for the users and the other for the blockage, over the line segment $[-V,V]$. The density of users is $\lambda$ and the density of blockage is $\mu$. Every user transmits with probability $\xi$, independently of other users and of its own transmissions in previous time slots. We use a bounded distance-based propagation pathloss model,  $l(r)\!=\!\min\left\{1,r^{-a}\right\}$, where $r$ is the distance and $a\geq 2$ is the pathloss exponent. In order to make the analysis valid also for sub-$6$ GHz cellular networks, we model the fast fading by the Rayleigh distrbution with unit mean. Also, there is correlated slow fading due to  blockage. The locations of obstacles are fixed but unknown. The obstacles do not hinder the user moves but they attenuate the user signal. It is assumed that the penetration loss per obstacle is uniformly distributed on $[0,\gamma],\gamma\leq 1$. 

Assuming common transmit power level $P_t$ for all users, the interference at time slot $t$ and location $y_p\in [-V,V]$ is 
\[
\mathcal{I}(t) = P_t \sum\nolimits_{i=1}^k {\xi_i(t)\, h_i(t) \, \beta_i(t)\, l\!\left(x_i(t)\!-\!y_{p}\right)}
\]
where $k$ is a particular realization of the PPP governing the distribution of users, $\xi_i$ is a Bernoulli \ac{RV} describing the $i$-th user activity, $\mathbb{E}\left\{\xi_i\right\}\!=\!\xi\, \forall i$, $h_i$ is an exponential \ac{RV} with unit mean modeling Rayleigh fast fading, $\mathbb{E}\left\{h_i\right\}\!=\!1\, \forall i$, $\beta_i$ is the \ac{RV} describing the penetration loss between the $i$-th user and the location $y_{p}$, and $x_i\in [-V,V]$ is a uniform \ac{RV} modeling the location for the $i$-th user.

The distribution of $\beta_i$ is difficult to obtain in terms of simple functions, however the moments of the penetration loss at distance $d_i\!=\!|x_i-y_p|$, i.e., between the $i$-th user and the location $y_p$ can be computed as $\mathbb{E}\!\!\left\{\beta_i^s\right\} = e^{-\mu d_i \left(1 - \frac{1}{1+s}\gamma^s \right)}$~\cite{Bai2014,Koufos2016b}.
%% \begin{equation}
%% \label{eq:Penetration}
%% \mathbb{E}\!\!\left\{\beta_i^s\right\} = e^{-\mu d_i \left(1 - \frac{\gamma^s}{1+s} \right)}. 
%% \end{equation}
Even though the users are distributed independently of each other, they may be blocked by some common obstacles. The first-order cross-moment of penetration loss for two users $i,j$ depends on the relative locations of $x_i,x_j$ w.r.t. $y_p$. When the two links $x_i\!\rightarrow\! y_p$ and $x_j\!\rightarrow\! y_p$ do not share any obstacles, the penetration losses are uncorrelated,  $\mathbb{E}\!\!\left\{\beta_i\beta_j\right\}\!=\!e^{-\mu(d_i+d_j)\left(1-\frac{\gamma}{2}\right)}$. Otherwise,  
$\mathbb{E}\!\!\left\{\beta_i\beta_j\right\} = e^{-\mu \min\left\{d_i,d_j\right\}\left( 1-\frac{1}{3}\gamma^2\right)} e^{-\mu |d_i-d_j| \left( 1-\frac{1}{2}\gamma\right)}$~\cite{Koufos2016b}.

In what follows, we will make use of the \ac{MGF} to analyze the moments of  interference. The \ac{MGF} of interference at time slots $t,\tau$ is 
%\[ \Phi_{\mathcal{I}} \!= \!\iiint\limits_{\rm h\, \boldsymbol \beta \, x}\sum_{{ {\rm \xi},i}}{e^{s_{\!1} \mathcal{I}(t) + s_2\mathcal{I}(\tau)}f_{\!\rm x,\boldsymbol \beta} \,f_{\xi}\, f_{{\rm h}} \, {\text{Po}}(\lambda) \, {\rm d x dh d} \boldsymbol\beta} \]
%\[ \Phi_{\mathcal{I}} \!= \!\int\nolimits_{\rm h}\!\int\nolimits_{\rm \boldsymbol \beta}\!\int\nolimits_{\rm x}\!\sum_{{ {\rm \xi},i}}{e^{s_{\!1} \mathcal{I}(t) + s_2\mathcal{I}(\tau)}f_{\!\rm x,\boldsymbol \beta} \,f_{\xi}\, f_{{\rm h}} \, {\text{Po}}(\lambda) \, {\rm d x dh d} \boldsymbol\beta} \] 
\[
\Phi_{\mathcal{I}} \!= \!\int\nolimits\!\!\! \int\nolimits\!\!\!\int\nolimits\!\!\sum\nolimits_{{ {\rm \xi},k}}{e^{s_{\!1} \mathcal{I}(t) + s_2\mathcal{I}(\tau)}f_{\!\rm x,\boldsymbol \beta} \,f_{\xi}\, f_{{\rm h}} \, {\text{Po}}(\lambda) \, {\rm d x \, dh \, d} \boldsymbol\beta}
\] 
where $\rm \xi$, $\rm h$, $\rm x$ and $\boldsymbol \beta$ are vectors of \acp{RV} with elements, $\xi_i, h_i$, $x_i$ and $\beta_i \, \forall i$ at time slots $t,\tau$, ${\text{Po}}(\lambda)\!=\!\frac{e^{-\lambda} \lambda^k}{k!}$ stands for the Poisson distribution, and the arguments in the Probability Distribution Functions  are omitted for brevity. 

In order to assess the correlation of interference at time slots $t,\tau$ we use the Pearson correlation coefficient, i.e., the ratio of the covariance of RVs $\mathcal{I}(t), \mathcal{I}(\tau)$ divided by the product of their standard deviations. We consider static users $\left(u\!=\!0\right)$, and users with infinite velocity $\left(u\!\rightarrow\!\infty\right)$. In the former, the locations of users are fixed but unknown. In the latter, a new realization of users is drawn in every time slot. In both cases, the statistics of interference are independent of the time slots $t,\tau$ we take the measurements and the time-lag $|t\!-\!\tau|$. Therefore the Pearson correlation coefficient can be written as %at time slots $t,\tau$ is independent of $t,\tau$ 
%Therefore the temporal correlation of interference becomes independent of the time-lag $l\!=\!|t\!-\!\tau|$, and the Pearson correlation coefficient at time slots $t,\tau$ is 
\begin{equation}
\rho = \frac{\mathbb{E}\!\left\{\mathcal{I}(t)\mathcal{I}(\tau)\right\}-\mathbb{E}\!\left\{\mathcal{I}(t)\right\}^2} {\mathbb{E}\!\left\{\mathcal{I}^2(t)\right\}-\mathbb{E}\!\left\{\mathcal{I}(t)\right\}^2}.
\label{eq:rho}
\end{equation}
%is independent of $t,\tau$. 

For the static case, we denote the correlation coefficient by $\rho_0$. For the mobile case, we denote it by $\rho_\infty$. The correlation coefficient is location-dependent but we omit the related index for brevity. We will show how to calculate the coefficients $\rho_0,\rho_\infty$ at the origin. The expressions at an arbitrary point $y_p\in [-V,V]$ can be obtained in a similar manner. 

\section{Interference mean and variance} 
\label{sec:Moments}
The mean of interference is computed after evaluating the first derivative of the \ac{MGF} $\frac{\partial \Phi_{\mathcal{I}}}{\partial s_{\!1}}$ at $s_1\!=\!0$. %\big|_{s_{\!1}=0}
%% \begin{subequations}
%% %\label{eq:Mean}
%% \arraycolsep=0.4pt\def\arraystretch{2.2}
%% \begin{align}%{lcl}
%% \mathbb{E}\!\left\{\mathcal{I}\right\} \stackrel{(a)}{=}& \!\sum\limits_i\!\mathbb{E}\!\left\{h_i\right\} \!\mathbb{E}\!\left\{\xi_i\right\} \!\!\!\int\nolimits_{\!x_i}\!\! \int\nolimits_{{\!\beta_i}}{\!\!\!\beta_i l\!\left(x_{\!i}\right) \!f_{\beta_{\!i}|x_{\!i}} f_{x_{\!i}} {\rm d}\beta_{\!i} {\rm d}x_i} {\text{Po}(\!\lambda\!)} \\ 
%% \stackrel{(b)}{=}& \displaystyle \lambda\xi \int_{-V}^V{\!\!\mathbb{E}\!\left\{\beta_x\right\} l(x){\rm d}x} \\ 
%% =& 2\lambda\xi\!\! \left(\! \frac{2\left(1\!-\!e^{-\!\frac{\mu\left(2-\gamma\right)}{2}}\right)}{\mu\left(2\!-\!\gamma\right)} \!+\!  E_{\!a}\!\left(\!\frac{\mu(2\!-\!\gamma)}{2}\!\right) \!-\!  
%% \frac{E_{\!a}\!\left(\!\frac{\mu(2\!-\!\gamma)V}{2}\!\right)}{V^{a-1}}\!\right)
%% \end{align}
%% \end{subequations}
\begin{equation}
\label{eq:Mean}
\arraycolsep=0.4pt\def\arraystretch{2.2}
\begin{array}{lcl}
\mathbb{E}\!\left\{\mathcal{I}\right\} &\stackrel{(a)}{=}& \displaystyle \sum\nolimits_i\!\mathbb{E}\!\left\{h_i\right\} \!\mathbb{E}\!\left\{\xi_i\right\} \!\!\int\nolimits\!\!\! \int\nolimits{\!\!\!\beta_i l\!\left(x_{\!i}\right) \!f_{\beta_{\!i}|x_{\!i}} f_{x_{\!i}} {\rm d}\beta_{\!i}\, {\rm d}x_i} \, {\text{Po}(\!\lambda\!)} \\ 
%&\stackrel{(b)}{=}& \displaystyle \lambda\xi \int_{-V}^V{\!\!\mathbb{E}\!\left\{\beta_x\right\} l(x){\rm d}x} \\ 
&\stackrel{(b)}{=}& 2\lambda\xi\!\! \left(\! \frac{2\left(1\!-\!e^{-\!\frac{\mu\left(2-\gamma\right)}{2}}\right)}{\mu\left(2\!-\!\gamma\right)} \!+\!  E_{\!a}\!\left(\!\frac{\mu(2\!-\!\gamma)}{2}\!\right) \!-\!  
\frac{E_{\!a}\!\left(\!\frac{\mu(2\!-\!\gamma)V}{2}\!\right)}{V^{a-1}}\!\right)
\end{array}
\end{equation}
where (a) follows from the fact that the penetration loss depends on the user location, (b) uses that the users are indistinct and also averages over the Poisson distribution ${\text{Po}(\!\lambda\!)}$,  $E_n(z)\!=\!\int_1^\infty{t^{-n}e^{-zt}{\rm d}t}$ is the generalized exponential integral, and the transmit power level has been taken equal to $P_t\!=\!1$. 
%\!=\!\int_1^\infty{\frac{e^{\!zt}}{t^n}{\rm d}t}

The second moment of interference is 
\begin{equation}
\label{eq:SecondMoment}
\arraycolsep=0.4pt\def\arraystretch{2.2}
\begin{array}{lcl}
 \mathbb{E}\!\left\{\mathcal{I}^2\right\}&=& \displaystyle 2 \lambda\xi \int_{-V}^V{\!\!\mathbb{E}\!\left\{\beta_x^2\right\} l^2(x){\rm d}x} \!+\! \sigma \\ 
& = &  4\lambda\xi\!\Bigg(\!\!\frac{3}{\mu\left(\!3\!-\!\gamma^2\!\right)}\left(\!1\!\!-\!\!e^{-\!\frac{1}{3}\mu\left(3\!-\!\gamma^2\right)}\!\right) \!+\!  E_{2a}\!\left(\!\frac{\mu}{3}(3\!-\!\gamma^2)\!\right) \!-\! \\ 
& & \,\,\, \frac{1}{V^{2a-1}}E_{2a}\!\left(\frac{1}{3}\mu(3\!-\!\gamma^2)V\right)\!\!\Bigg) \!+\!\sigma 
\end{array}
\end{equation}
where it has been used that $\mathbb{E}\!\left\{h^2\right\}\!=\!2, \mathbb{E}\!\left\{\xi^2\right\}\!=\!\xi$, and the term $\sigma$ captures the correlation in the interference levels generated by different users 
\begin{equation}
\sigma\!=\lambda^2\xi^2\!\int_{-\!V}^V\int_{-\!V}^V{\!\!\mathbb{E}\left\{\beta_x\beta_y\right\}l(x)l(y){\rm d}y \, {\rm d}x}.
\label{eq:sigma}
\end{equation}

The calculation of $\sigma$ can be split into two terms, $\sigma\!=\!\sigma_1\!+\!\sigma_2$, depending on whether pairs of links share common obstacles or not. The uncorrelated part is equal to $\sigma_1\!=\! \frac{1}{2}\mathbb{E}\left\{\mathcal{I}\right\}^2$, and the correlated part can be written as $\sigma_2\!=\!4 \lambda^2\xi^2\!\int_0^V\!\!\int_0^x{\!\mathbb{E}\left\{\beta_x\beta_y\right\}l(x)l(y){\rm d}y \, {\rm d}x}$. In order to calculate $\sigma_2$, one has to take care of the piecewise nature of the pathloss model. For a positive $\gamma$, we finally get
%\[
%\begin{equation}
%\sigma_2\!=\!4 \lambda^2\xi^2\!\int_0^V\!\!\int_0^x{\!\!\mathbb{E}\left\{\beta_x\beta_y\right\}l(x)l(y){\rm d}y \, {\rm d}x}.
%\label{eq:Sigma2}
%\end{equation}
%\]
\begin{equation}
\label{eq:sigma2}
\arraycolsep=0.4pt\def\arraystretch{2.2}
\begin{array}{lcl}
\sigma_2 &=& 4 \lambda^2\xi^2\! \Bigg (\! 6\frac{3(2\!-\!\gamma)e^{-\!\frac{1}{3}\mu\left(3\!-\gamma^2\right)}\!+\!\gamma\left(3\!-\!2\gamma\right)\!-\!2(3\!-\!\gamma^2) e^{-\!\frac{1}{2}\mu(2-\gamma)}} {\mu\gamma^2(2-\gamma)(3-2\gamma)(3-\gamma^2)} +  \\ 
& & \frac{6\left(2\!-\!\gamma\right) \left(\! 1-e^{-\frac{1}{6}\mu\gamma\left(3-2\gamma\right)}\!\right) \left(\! E_a\!\left(\!\frac{\mu(2\!-\!\gamma)}{2}\!\right) - E_a\!\left(\!\frac{\mu(2\!-\!\gamma)V}{2}\!\right) \!\right) }{ \mu\gamma\left(6-7\gamma + 2\gamma^2\right)} + \\
& & E_a\!\left(\!\frac{\mu\gamma\left(3\!-\!2\gamma\right)}{6}\!\right)\!\left( E_a\!\left(\!\frac{\mu(2\!-\!\gamma)}{2}\!\right) \!-\!V^{1\!-\!a} \!E_a\!\left(\!\frac{\mu(2\!-\!\gamma)V}{2}\!\right)\right)  + \\
& & \displaystyle \int_1^V{\!\!\!e^{-\!\frac{1}{2}\mu\left(2-\gamma\right)x} x^{1\!-\!2a}E_a\!\!\left(\!\frac{1}{6}\mu\gamma\left(3\!-\!2\gamma\right)x\!\right)\!\!{\rm d}x} \Bigg ).
\end{array}
\end{equation}

In equation~\eqref{eq:sigma2}, the integral $I_0\!=\!\int_1^V{e^{-cx}x^{1-2\alpha}E_a\!\!\left(b\, x\right){\rm d}x}$, where $c\!=\!\frac{\mu(2-\gamma)}{2}$ and $b\!=\!\frac{\mu\gamma(3-2\gamma)}{6}$ has the least contribution of the four terms. It can be computed in terms of the incomplete Gamma function only if the constants $c,b$ are equal. This is not true unless $\mu\!=\!0$, where the integral becomes trivial to solve and equals  $\frac{1-V^{2(1-a)}}{2(a-1)^2}$. For a positive $\mu$, the integral decays sharply with $x$. One may avoid numerical integration, and use the Laplace method to approximate it instead. Due to the lack of space, we give  only the second-order approximation for $V\!\rightarrow\!\infty$, $I_0 \approx e^{-c+\log\!E_a\!\left(b\right)} \!\left(\frac{1}{A} - \frac{2B}{A^3} \right)$, where $A\!=\!2a\!-\!1\!+\!c\!-\!\frac{b E_{a-1}(b)}{E_a(b)}$,  $B\!=\!\frac{b^2E_{a-2}(b)}{2E_a(b)} -\frac{b^2 E_{a-1}(b)^2}{2E_a(b)^2}- \frac{2a-1}{2}$. Even this  has sufficient accuracy for our problem.   
%The approximation becomes more accurate for increasing density of blockage. 
%% \[
%% I_0 \approx e^{-c+\log\!E_a\!\left(b\right)} \!\left(\frac{1}{A} - \frac{2B}{A^3} \right)
%% %e^{-c+\log\!E_a\!\left(b\right)} \!\left(\frac{1-e^{-\!A(1-V)}}{A} + \frac{B\left(2-2e^{-A(1-V)}\left(2-A(1-V)(2-A(1-V)\right)\right)}{A^3} \right)
%% \]

%% \begin{equation}
%% \label{eq:sigma2}
%% \arraycolsep=0.4pt\def\arraystretch{2.2}
%% \begin{array}{lcl}
%% \sigma_2 &=& 2 \lambda^2\xi^2\! \Bigg (\! 6\frac{3(2\!-\!\gamma)e^{-\!\frac{1}{3}\mu\left(3\!-\gamma^2\right)}\!+\!\gamma\left(3\!-\!2\gamma\right)\!-\!2(3\!-\!\gamma^2) e^{-\!\frac{1}{2}\mu(2-\gamma)}} {\mu\gamma^2(2-\gamma)(3-2\gamma)(3-\gamma^2)} +  \\ 
%% & & \frac{6\!\left(\!1\!-\!e^{-\!\frac{1}{6}\mu\gamma\left(3-2\gamma\right)}\!\right)\! E_a\!\left(\!\frac{2\mu\!-\!\mu\gamma}{2}\!\right)}{\mu\gamma(3-2\gamma)} \!+\!\! E_a\!\!\left(\!\frac{\mu\gamma\left(3\!-\!2\gamma\right)}{6}\!\right)\!\! E_a\!\!\left(\!\frac{2\mu\!-\!\mu\gamma}{2}\!\right) \!+\! \\
%% & & \displaystyle \int_1^\infty{\!\!\!e^{-\!\frac{1}{2}\mu\left(2-\gamma\right)x} x^{1\!-\!2a}E_a\!\!\left(\!\frac{1}{6}\mu\gamma\left(3\!-\!2\gamma\right)\!\right)\!\!{\rm d}x} \Bigg ).
%% \end{array}
%% \end{equation}

For impenetrable blockage, one has to substitute  $\gamma\!=\!0$ in equations~\eqref{eq:Mean},~\eqref{eq:SecondMoment}. For $\gamma\!=\!0$, equation~\eqref{eq:sigma2} becomes indefinite. One should use $\sigma_2 = 4\lambda^2\xi^2\!\Big(\!\frac{1-(1\!+\!\mu)e^{-\!\mu}}{\mu^2}  \!+\! \frac{a}{a-1}\!\left(\!E_{\!a}\!\left(\mu\right) \!-\! \frac{E_{\!a}\left(\mu V\right)}{V^{a-1}}\right) + \frac{V^{2(1\!-a)}E_{2a-1}\!\left(\mu V\right)- E_{2a-1}\!\left(\mu\right)}{a-1} \Big)$ instead. 
%% \begin{equation}
%% \arraycolsep=0.4pt\def\arraystretch{2.2}
%% \begin{array}{lcl}
%% \sigma_2 &=& 4\lambda^2\xi^2\!\Bigg(\!\frac{1-(1\!+\!\mu)e^{-\!\mu}}{\mu^2}  \!+\! \frac{a}{a-1}\!\left(\!E_{\!a}\!\left(\mu\right) \!-\! \frac{E_{\!a}\left(\mu V\right)}{V^{a-1}}\right) + \\ & & \,\,\, \frac{\Gamma\!\left(2(1-a),\mu V \!\right)-\Gamma\!\left(2(1-a),\mu \!\right)}{\mu^{2(1-a)} (a-1)} \Bigg).
%% \end{array}
%% \end{equation}
%% \[
%% \sigma_2\!=\!2\lambda^2\xi^2\left(\frac{1\!-\!e^{-\!\mu}\!-\!\mu e^{-\!\mu}}{\mu^2}  \!+\! \frac{aE_{a}\!\!\left(\mu\right) - E_{2a\!-\!1}\!\!\left(\mu\right)}{a-1}\right).
%% \]

\section{Temporal interference correlation}
\label{sec:Correlation}
\begin{figure}[!t]
 \centering
  \includegraphics[width=3.5in]{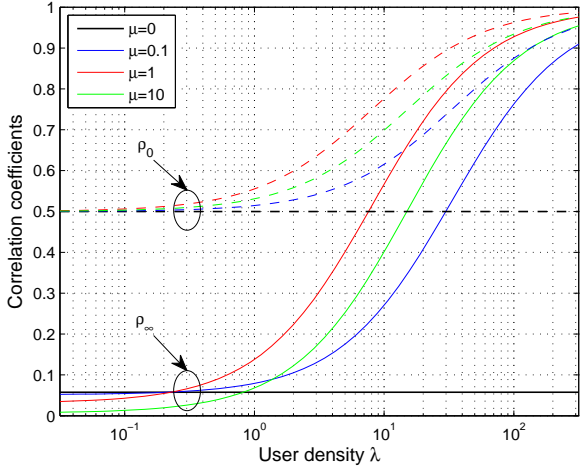}
 \caption{Correlation coefficients of interference $\rho_0, \rho_\infty$ w.r.t. the user density. Minimum penetration loss $\gamma\!=\!1$, pathloss exponent $a\!=\!2$, size of the deployment domain $V\!=\!25$ and continuous user activity $\xi\!=\!1$.}
 \label{fig:VariableLamda} 
\end{figure}
The cross-correlation of interference can be computed from the first-order cross-derivative of the \ac{MGF}, $\frac{\partial^2 \Phi_{\mathcal{I}}}{\partial s_{\!1}\partial s_{\!2}}$ at $(s_1,s_2)\!=\!(0,0)$. For the static case,  the penetration losses of a single user at different time slots are fully correlated. Hence,  %$s_1\!=\!0, s_2\!=\!0$. 
\begin{equation}
\label{eq:CrossStatic}
\mathbb{E}\!\left\{\mathcal{I}(t)\mathcal{I}(\tau)\right\} = \lambda\xi^2 I  + \sigma 
\end{equation}
where $I\!=\! \int_{-V}^V{\!\mathbb{E}\!\left\{\beta_x^2\right\} l^2(x){\rm d}x}$ is computed as in~\eqref{eq:SecondMoment}. 

With infinite velocity, the locations of a user at different time slots are uncorrelated but the penetration losses may still be correlated. Hence, 
\begin{equation}
\label{eq:CrossMobile}
\mathbb{E}\!\left\{\mathcal{I}(t)\mathcal{I}(\tau)\right\} = \frac{\lambda\xi^2}{2V} \!\!\! \int_{-V}^V\!\!\int_{-V}^V{\!\!\!\!\!\mathbb{E}\!\left\{\beta_x \beta_y\right\} l(x) l(y){\rm d}x \, {\rm d}y} + \sigma. 
\end{equation}

Using equation~\eqref{eq:sigma}, the first term in equation~\eqref{eq:CrossMobile} can also be written as $\frac{\sigma}{2\lambda V}$.  The correlation coefficients are computed after substituting equations~\eqref{eq:CrossStatic},~\eqref{eq:CrossMobile} in equation~\eqref{eq:rho}%can be written as
\begin{equation}
\label{eq:Coefficients}
\arraycolsep=0.4pt\def\arraystretch{2.2}
\begin{array}{lcl}
%% \rho_0 &=& \frac{\lambda\xi^2 I + \sigma - \mathbb{E}\left\{\mathcal{I}\right\}^2}{2\lambda\xi I + \sigma - \mathbb{E}\left\{\mathcal{I}\right\}^2} \\ 
%% \rho_\infty &=& \frac{\frac{\sigma}{2\lambda V} + \sigma - \mathbb{E}\left\{\mathcal{I}\right\}^2}{2\lambda\xi I + \sigma - \mathbb{E}\left\{\mathcal{I}\right\}^2}.
\rho_0 = \frac{\lambda\xi^2 I + \sigma - \mathbb{E}\left\{\mathcal{I}\right\}^2}{2\lambda\xi I + \sigma - \mathbb{E}\left\{\mathcal{I}\right\}^2} &,& 
\,\, \rho_\infty = \frac{\frac{\sigma}{2\lambda V} + \sigma - \mathbb{E}\left\{\mathcal{I}\right\}^2}{2\lambda\xi I + \sigma - \mathbb{E}\left\{\mathcal{I}\right\}^2}.
\end{array}
\end{equation}
%For infinite line, this term  vanishes to zero and $\mathbb{E}\!\left\{\mathcal{I}(t)\mathcal{I}(\tau)\right\}\!=\!\sigma$.
\subsubsection{No blockage, $\mu\!=\!0$} 
Without blockage, the interference levels generated by different users become uncorrelated, i.e., $\sigma\!=\!\mathbb{E}\!\left\{\mathcal{I}\right\}^2$. After substituting $\sigma\!=\!\mathbb{E}\!\left\{\mathcal{I}\right\}^2$ in equation~\eqref{eq:Coefficients}, and this back in~\eqref{eq:rho}, we get $\rho_{0}|_{\mu\!=\!0}\!=\!\frac{\xi}{2}$, and 
\begin{equation}
\label{eq:NoBlockageInf}
\rho_{\!\infty}|_{\mu\!=\!0}\!=\!\frac{\mathbb{E}\!\left\{\mathcal{I}\right\}^2}{4\lambda^2 \xi V I } \!=\!\frac{\xi\left(\!a\!-\!V^{1\!-\!a}\!\right)^2 (2a\!-\!1)}{2V\!(a\!-\!1)^2\left(2a\!-\!V^{1\!-\!2a}\right)}\!>\!0.
\end{equation}
%For a large $V$, the terms $V^{1\!-\!a}, V^{1\!-\!2a}$ in equation~\eqref{eq:NoBlockageInf} can be eliminated,   $\rho_\infty|_{\mu\!=\!0}\!\approx\! \frac{\xi\left(2a-1\right)a}{4V\left(a-1\right)^2}$. 
\begin{remark}
\label{remark:1}
A bounded domain introduces correlation in the distance-based propagation pathloss, and makes the interference level correlated in time, even if the user mobility is uncorrelated. For infinite line, $\lim_{V\!\rightarrow \! \infty}\rho_\infty|_{\mu\!=\!0}\!=\!0$. 
\end{remark}

\subsubsection{Blockage $\mu\!>\!0$}
With a finite blockage density  $\sigma\!\!>\!\!\mathbb{E}\!\!\left\{\mathcal{I}\right\}^{\!2}\!\!\!.$
%their limits at user density $\lambda\!\rightarrow\!\infty$ are equal to one. 
%After substituting in equation~\eqref{eq:Coefficients} the terms $\mathbb{E}\left\{\mathcal{I}\right\}$ and $\sigma$ from equations~\eqref{eq:Mean} and~\eqref{eq:sigma} respectively, and
\begin{remark}
\label{remark:2}
For infinite line, $\lim_{V\!\rightarrow\!\infty}\frac{\sigma}{2\lambda V}\!=\!0$. Hence, from equation~\eqref{eq:Coefficients},  $\lim\limits_{V\!\rightarrow\!\infty}\rho_\infty\!=\!\lim\limits_{V\!\rightarrow\!\infty}\frac{\sigma - \mathbb{E}\left\{\mathcal{I}\right\}^2}{2\lambda\xi I + \sigma - \mathbb{E}\left\{\mathcal{I}\right\}^2} \!>\!0$. Thus, the coefficient $\rho_\infty$, unlike the coefficient $\rho_\infty|_{\mu\!=\!0}$, is positive even if the deployment area is infinite. 
%As we will show next, $\rho_\infty\!\rightarrow\! 0$ only in the limit $\mu\!\rightarrow\!\infty$. 
%Based on Remark~\ref{remark:2}, equations~\eqref{eq:Mean} and~\eqref{eq:sigma}, one can see that for a fixed density of blockage, the limits of correlation coefficients $\rho_0, \rho_\infty$ at user density $\lambda\!\rightarrow\!\infty$ are equal to one.
\end{remark}
\begin{remark}
\label{remark:3}
Starting from equation~\eqref{eq:Coefficients} and using that $\xi\leq 1$, one can show that in a static network, blockage increases the correlation of interference, i.e., $\rho_0\geq\frac{\xi}{2}=\rho_0|_{\mu\!=\!0}$.
\end{remark}
\begin{remark}
\label{remark:4}
Using that the Pearson correlation coefficient is at most equal to one, one can show that the first derivative of $\rho_\infty$ in equation~\eqref{eq:Coefficients} w.r.t.  $\lambda$ is positive. Also,  $\lim\limits_{\lambda\!\rightarrow\!\infty}\rho_\infty\!=\!1\!>\!\frac{\xi}{2}\!=\!\rho_0|_{\mu\!=\!0}\!>\!\rho_\infty|_{\mu\!=\!0}$, and $\lim\limits_{\lambda\!\rightarrow\!0}\!\rho_\infty\!=\!\frac{\xi\int_{-\!V}^V\int_{-\!V}^V{\!\!\mathbb{E}\left\{\beta_x\beta_y\right\}l(x)l(y){\rm d}y \, {\rm d}x}}{4V\int_{-V}^V{\!\mathbb{E}\!\left\{\beta_x^2\right\} l^2(x){\rm d}x}}\!\leq\!\frac{\xi\int_{-\!V}^V\int_{-\!V}^V{\!\!l(x)l(y){\rm d}y \, {\rm d}x}}{4V\int_{-V}^V{\! l^2(x){\rm d}x}}\!=\!\rho_\infty|_{\mu\!=\!0}$. Therefore with uncorrelated mobility, blockage reduces the correlation of interference at low user densities, while the opposite is true at high user densities. There will be a critical user density $\lambda^*$ where $\rho_\infty^*\!=\!\rho_\infty|_{\mu\!=\!0}$.
%% Using that the Pearson correlation coefficient is at most equal to one, one can show that the first derivatives of $\rho_0,\rho_\infty$ in~\eqref{eq:Coefficients} w.r.t.  $\lambda$ are positive. The limits of the coefficients at a high and low user density are  $\lim_{\lambda\!\rightarrow\!\infty}\!\!\left\{\rho_0,\rho_\infty\right\}\!=\!1$,  $\lim\nolimits_{\lambda\!\rightarrow\!0}\rho_0\!=\!\frac{\xi}{2}$, and $\lim_{\lambda\!\rightarrow\!0}\!\rho_\infty\!\leq\!\frac{\xi}{4V}<\rho_\infty|_{\mu\!=\!0}$. Therefore with uncorrelated mobility, blockage reduces the correlation of interference at low user densities, while the opposite is true at high user densities. There will be a critical user density $\lambda^*$ where $\rho_\infty^*\!=\!\rho_\infty|_{\mu\!=\!0}$.
\end{remark}
Let us denote by $p\!=\!\frac{\lambda}{\mu}$ the ratio of user density to blockage density. If we expand the moments of  interference around $\mu\!\rightarrow\!\infty$, we get $\mathbb{E}\left\{\mathcal{I}\right\}\!=\!\frac{2 p \xi}{2-\gamma}$, $\sigma_2\!=\!\frac{24 p^2 \xi^2}{(2-\gamma)(3-\gamma^2)}$ and $\mathbb{E}\left\{\mathcal{I}^2\right\}\!=\!\frac{12 p \xi}{3-\gamma^2}+\sigma$. After substituting these approximations in equation~\eqref{eq:Coefficients}, the correlation coefficients $\rho_0$ and $\rho_\infty$ around $\mu\!\rightarrow\!\infty$, keeping $p$ finite or $p\!\rightarrow\! 0$, can be read as  
\begin{equation}
\label{eq:Taylor}
\arraycolsep=0.1pt\def\arraystretch{2.2}
\begin{array}{lll}
\rho_0|_{\mu\!\rightarrow\!\infty} & = &\frac{3\xi\left(2\!-\!\gamma\right)^2\!+\! 12\xi p\left(2\!-\!\gamma\right) -4\xi p\left(3\!-\!\gamma^2\right)}{6\left(2\!-\!\gamma\right)^2\!+\! 12\xi p\left(2\!-\!\gamma\right) -4\xi p\left(3\!-\!\gamma^2\right)} \!=\! \frac{\xi}{2} +\! \\ 
&&  \frac{\left(\!3\!-\!3\gamma\!+\!\gamma^2\!\right) \!\left(2\!-\xi\right) \xi p}{3\left(2-\gamma\right)^2} \!-\!\!\frac{2\left(\!3\!-\!3\gamma\!+\!\gamma^2\!\right)^{\!2} \!\left(2-\xi\right) \xi^2 \!p^2}{9\left(2-\gamma\right)^4} \!+\! \mathcal{O}\!\left(p\right)^{\!3} \\ 
\rho_\infty|_{\mu\!\rightarrow\!\infty} & = &\frac{\left(1+\frac{1}{2\lambda V}\right) 6\xi p\left(2\!-\!\gamma\right) \!-\!\left(\frac{1}{2}-\frac{1}{2\lambda V}\right) 4\xi p\left(3\!-\!\gamma^2\right)}{3\left(2\!-\!\gamma\right)^2\!+\!6\xi p\left(2\!-\!\gamma\right) \!-\!2\xi p\left(3\!-\!\gamma^2\right)} \!=\! \frac{3\xi p}{2-\gamma} -\! \\ 
& & \frac{2\xi\left(3-\gamma^2\right)p} {3\left(2-\gamma\right)^2} \!-\! \frac{\left(6\xi\left(2-\gamma\right)-2\xi\left(3-\gamma^2\right)\right)^2\!p^2} {9\left(2-\gamma\right)^4} \!+\! \mathcal{O}\!\left(p\right)^3 
\end{array}
\end{equation}
where in the expression of $\rho_\infty|_{\mu\!\rightarrow\!\infty}$, the contribution of the terms $\frac{1}{2\lambda V}$ has been omitted from the series expansion of the fraction. This would be a valid approximation for a large $V$. 
\begin{remark}
\label{remark:5}
At a high density of blockage, the correlation coefficients increase with the user-to-blockage density ratio. 
%decrease with the blockage density $\mu$.
\end{remark}
%% few remarks can be drawn based on equation~\eqref{eq:Taylor}. 
%% \begin{remark} 
%% \label{remark:3}
%% (i)    (ii) The coefficient $\rho_\infty|_{\mu\!\rightarrow\!\infty}$ decreases with the size of the deployment domain $V$, while the coefficient $\rho_0|_{\mu\!\rightarrow\!\infty}$ is insensitive to $V$.  
%% \end{remark}
%(ii)  $\rho_0|_{\mu\!\rightarrow\!\infty}\geq\frac{\xi}{2}$, while $\rho_\infty|_{\mu\!\rightarrow\!\infty}$ may be larger or smaller than $\rho_\infty|_{\mu\!=\!0}$.
%Actually, for a finite user density $\lambda$ and $\mu\!\rightarrow\!\infty$, or equivalently $p\!\rightarrow\! 0$, the coefficients become equal to theconverge to $\rho_0\!\rightarrow\!\frac{\xi}{2}$ and $\rho_\infty\!\rightarrow\! 0$.

In Fig.~\ref{fig:VariableLamda}, we have used equation~\eqref{eq:Coefficients} to compute the correlation coefficients $\rho_0, \rho_\infty$ for various user and blockage densities. In the static case, blockage makes the propagation pathloss of different users correlated resulting in higher correlation coefficients than in the case without blockage, see Remark~\ref{remark:3}. In the mobile case, the impact of blockage on the interference correlation depends on the user density, see Remark~\ref{remark:4}: When the user density is low, the interference level is also low, and it would vary significantly with  mobility because of the transitions in the propagation conditions, from \ac{LoS} to \ac{NLoS} and vice versa. These transitions make the correlation of interference less than in the case without blockage. On the other hand, when the user density is high, the correlation of penetration losses among the user prevails, and mobility does not help much in reducing it. Some users will transit from \ac{LoS} to \ac{NLoS} but at the same time, some others with transit from \ac{NLoS} to \ac{LoS}. Overall, the interference level will not vary much.  %the temporal correlation of interference. 
%When the user density is low, the correlation of penetration losses among the users is also low, and at the same time there are transitions in the propagation conditions due to mobility, from \ac{LoS} to \ac{NLoS} and vice versa. These transitions make the correlation of interference less than in the case without blockage. On the other hand, when the user density is high, the correlation of penetration losses among the user prevails, and mobility does not help much in reducing it.
%coefficient $\rho_\infty$ could be well-approximated with the coefficient $\rho_\infty|_{\mu\!\rightarrow\!\infty}$.
When $\mu\!=\!10$, the approximations for a high density of blockage, see  equation~\eqref{eq:Taylor} become valid.  For the parameter settings used to generate Fig.~\ref{fig:VariableLamda}, $\gamma\!=\!1, \xi\!=\!1$, we get $\rho_\infty|_{\mu\!\rightarrow\!\infty} \!\approx\!\frac{2p}{3+2p}$ after neglecting the contribution of the term $\frac{1}{2\lambda V}$. From equation~\eqref{eq:NoBlockageInf} we get  $\rho_\infty|_{\mu\!=\!0}\!\approx\!\frac{3}{2V}$ for $a\!=\!2$ and after neglecting the contribution of the terms $V^{1-a},V^{1-2a}$. Therefore, $\rho_\infty|_{\mu\!\rightarrow\!\infty}\geq\rho_\infty|_{\mu\!=\!0}$ for $\lambda\!\geq\!\lambda^*, \lambda^*\!=\!\frac{9\mu}{4V\!-6}\!\approx\!0.95$, see Fig.~\ref{fig:VariableLamda}. To sum up, for a high density of blockage, the critical user-to-blockage density ratio can be expressed in a closed-form in terms of  the size of the deployment area $V$, the channel model $a,\gamma$ and the user activity $\xi$.
%In a similar manner, one can solve the inequality $\rho_\infty\!\geq\!\rho_\infty|_{\mu\!=\!0}$ and obtain the critical density of users $\lambda^*$ for an arbitrary density of blockage. 
%as can also be seen in the figure. 
%Note that for a high density of obstacles, e.g., 

\begin{figure}[!t]
 \centering
  \includegraphics[width=3.5in]{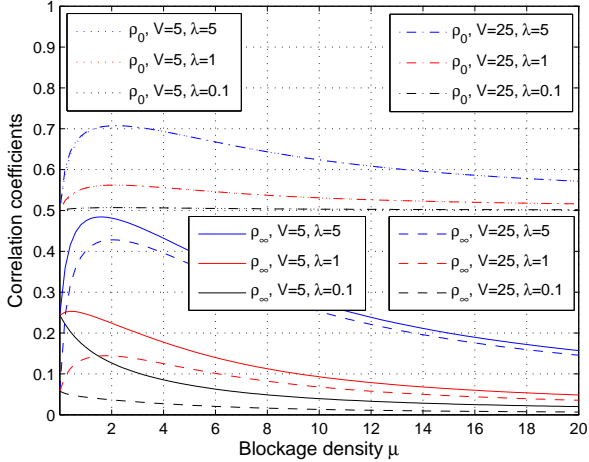}
 \caption{Correlation coefficients of interference $\rho_0,\rho_\infty$ w.r.t. the blockage density. The  parameter settings are available in the caption of Fig.~\ref{fig:VariableLamda}, unless otherwise stated in the legend.}
 \label{fig:VariableMu}
\end{figure}

%The interplay between blockage density, user density and interference correlation is also depicted in Fig.~\ref{fig:VariableMu}. 
When the user density is fixed and finite and the blockage density keeps on increasing, the correlation of penetration losses from different users starts to reduce beyond a certain density of blockage. As a result, the correlation coefficients $\rho_0,\rho_\infty$ will reduce too, see Fig.~\ref{fig:VariableMu} and Remark~\ref{remark:5}. 
%This behavior can also be deduced from equation~\eqref{eq:Taylor}. 
%This is also shown in equation~\eqref{eq:Taylor}. 
In Fig.~\ref{fig:VariableMu}, we also see that smaller domains $V$ are associated with higher correlation coefficients $\rho_\infty$. This is because a smaller domain results in less randomness in the distance-based propagation pathloss of a user at different time slots. Obviously, the impact of distance-based pathloss on the interference is more prominent at low blockage densities. In the static case, the size of the deployment domain does not impact much the correlation of interference. The curves for different domains $V$ in Fig.~\ref{fig:VariableMu} practically overlap. %The randomness in the interference level is mostly determined from the densities of users and blockage.

%Thus far, we have studied the system behavior only at the origin. 
To get a glimpse on the location-dependent properties of interference correlation, we also study it at the boundary, $y_p\!=\!V$. Without blockage, the correlations coefficients are  $\rho_{0}|_{\mu\!=\!0}\!=\!\frac{\xi}{2}$ and $\rho_{\!\infty}|_{\mu\!=\!0}\!\approx\! \frac{\xi\left(\!a\!-\!V^{1\!-\!a}\!\right)^2 (2a\!-\!1)}{4V\!(a\!-\!1)^2\left(2a\!-\!V^{1\!-\!2a}\right)}$. %$\frac{\xi\left(2a-1\right)a}{8V\left(a-1\right)^2}$.  
%The calculations carried out in Section~\ref{sec:Moments} and Section~\ref{sec:Correlation} could be extended at an arbitrary point $y_p\in [-V,V]$. %$\rho_{\!\infty}|_{\mu\!=\!0}\!=\!\frac{\xi\left(\!a-\left(2V\right)^{1\!-\!a}\!\right)^2 (2a-1)}{4V\!(a-1)^2\left(2a-\left(2V\right)^{1\!-\!2a}\right)}$. For a large $V$, we get 
\begin{remark}
\label{remark:6}
The coefficient $\rho_{\!\infty}|_{\mu\!=\!0}$ at the boundary is half the coefficient $\rho_{\!\infty}|_{\mu\!=\!0}$ at the center because at the boundary there is more randomness in the distance-based pathloss. 
%The coefficient $\rho_{0}|_{\mu\!=\!0}$ does not depend on the location. 
\end{remark}

\begin{figure}[!t]
 \centering
  \includegraphics[width=3.5in]{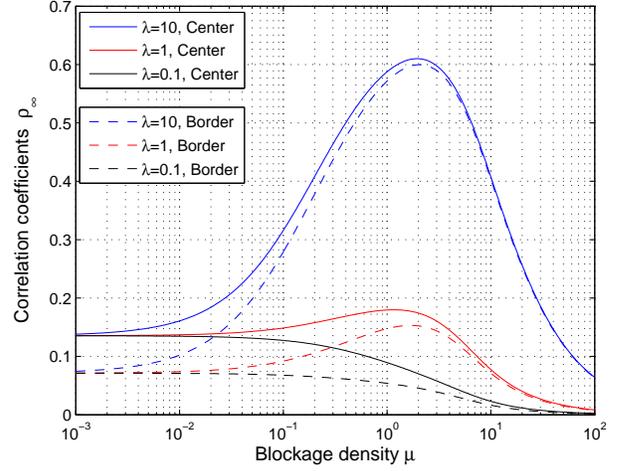}
 \caption{Correlation coefficients $\rho_\infty$ at the origin and at the boundary w.r.t. the blockage density. The size of the domain is $V\!=\!10$. The rest of parameter settings are available in the caption of Fig.~\ref{fig:VariableLamda}.}
 \label{fig:Border}
\end{figure}

With blockage, the coefficient $\rho_0$ at the boundary will be marginally higher than the coefficient $\rho_0$ at the center, because the boundary sees more correlated penetration losses. On the other hand, the  coefficient $\rho_\infty$ is smaller at the boundary than at the center, see Fig.~\ref{fig:Border}. This is because at the boundary, where the level of interference is also less, the randomness in the distance-based propagation pathloss is higher. For increasing density of blockage, the generated interference is dominated from the users located close to the boundary. Therefore the higher randomness of the link gains starts to vanish and the correlation becomes less sensitive to the location, see Fig.~\ref{fig:Border}. It can be shown that for a high density of blockage, the coefficient $\rho_\infty$ at the boundary can also be approximated by the expression in equation~\eqref{eq:Taylor}. %used to approximate the coefficient  $\rho_\infty$ at the center, see equation~\eqref{eq:Taylor}.

\section{Conclusions}
In this letter, we showed that a bounded domain and/or a domain with blockage  can induce temporal correlation of interference even if the user locations are uncorrelated over time. With blockage, the correlation coefficient increases with the density of users. Therefore beamforming techniques, which essentially scale down the density of users generating interference, will scale down the temporal correlation of interference too. Extending the results of this paper in two-dimensional areas with beamforming and nonuniform distribution of users, e.g., due to \ac{RWPM} mobility is a topic for future work.  
% of a user at different time slots
%However, in Fig.~\ref{fig:Border} we also see that for increasing blockage density, the coefficient $\rho_\infty$ becomes less and less sensitive to the location. For a high density of blockage, the generated interference is dominated from the users located close to the location $y_p$. Therefore the higher randomness of the link gains of a user at different time slots starts to vanish for increasing density of blockage. It can be shown that for a high density of blockage, the coefficient $\rho_\infty$ at the boundary can also be approximated by the expression used to approximate the coefficient  $\rho_\infty$ at the center, see equation~\eqref{eq:Taylor}. 

\end{document}